\documentclass{article}
\usepackage{spconf,amsmath,graphicx}

\usepackage{caption} 
\usepackage{tabularx}
\usepackage{float}
\usepackage{array}  
\usepackage{booktabs} 
\usepackage{multirow}
\usepackage{makecell}

\usepackage{color}

\usepackage{bbding}

\usepackage{amsfonts,amssymb} 

\usepackage{subfigure} 

\title{Multi-level graph learning for audio event classification and human-perceived annoyance rating prediction}

\name{Yuanbo Hou$^{1}$, Qiaoqiao Ren$^1$, Siyang Song$^2$, Yuxin Song$^3$, Wenwu Wang$^4$, Dick Botteldooren$^{1}$}
 \address{
 $^1$Ghent University, Belgium $^2$University of Leicester, UK  $^3$Baidu Inc., China $^4$University of Surrey, UK }

\begin{document}

\captionsetup{font={small}}

\maketitle
\begin{abstract}

\vspace{-0.1cm}
\noindent
WHO's report on environmental noise estimates that 22 M people suffer from chronic annoyance related to noise caused by audio events (AEs) from various sources. Annoyance may lead to health issues and adverse effects on metabolic and cognitive systems. 
In cities, monitoring noise levels does not provide insights into noticeable AEs, let alone their relations to annoyance. 
To create annoyance-related monitoring, this paper proposes a graph-based model to identify AEs in a soundscape, and explore relations between diverse AEs and human-perceived annoyance rating (AR). Specifically, this paper proposes a lightweight multi-level graph learning (MLGL) based on local and global semantic graphs to simultaneously perform audio event classification (AEC) and human annoyance rating prediction (ARP). Experiments show that: 
1) MLGL with 4.1 M parameters improves AEC and ARP results by using semantic node information in local and global context-aware graphs;
2) MLGL captures relations between coarse- and fine-grained AEs and AR well; 3) Statistical analysis of MLGL results shows that some AEs from different sources significantly correlate with AR, which is consistent with previous research on human perception of these sound sources.

\end{abstract}

\vspace{-0.1cm}

\vspace{-0.4cm}
\section{Introduction}
\label{sec:intro}

\vspace{-0.3cm}
World Health Organization (WHO) reports that environmental noise is the 2nd cause of ill health in Europe after air pollution \cite{WHO}. 
WHO hereby considers that severe annoyance can lead to cardiovascular disease~\cite{munzel2018environmental} and reduced quality of life, with annoyance being an important pathway between noise and long-term health effects. Noticed audio events (AEs) may play a vital role in the emergence of annoyance~\cite{de2009model}. Although most epidemiological studies guided by \cite{world2018environmental} consider a single type of sound source, different sounds occur together in real life, and a person's annoyance rating (AR) will be affected by the combined effect of sound sources. Thus, this paper links audio event classification (AEC)~\cite{AEC} and annoyance rating prediction (ARP)~\cite{HGRL}. AEC aims to determine whether there are predefined AEs in the soundscape, while ARP aims to predict the overall human-perceived AR.
We are aware of the large gap between human AR in different environments, but we see our efforts as a first step towards deploying long-term monitoring of adverse effects of environmental sound.

\label{ssec:figure-f}
\begin{figure*}[t] 
	\setlength{\abovecaptionskip}{-0cm}  
	\setlength{\belowcaptionskip}{-0.4cm}   
	\centerline{\includegraphics[width = 0.9 \textwidth]{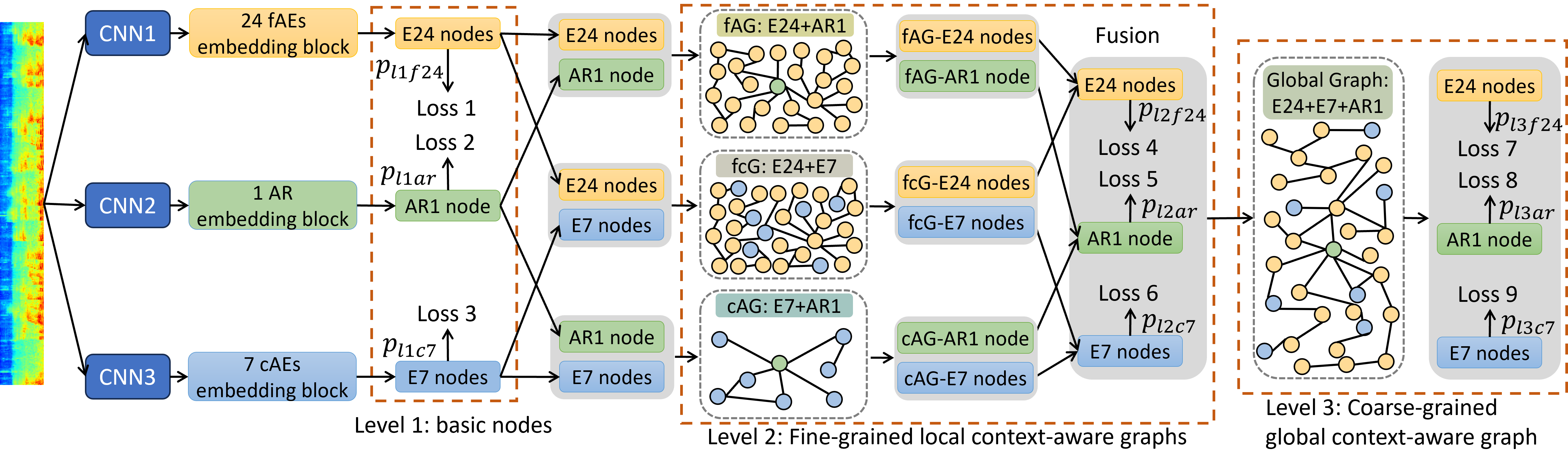}}
	\caption{The proposed lightweight attention-fused multi-level graph learning (MLGL).}
	\label{model}
\end{figure*}

\vspace{-0.1cm}
Most deep learning-based AEC-related studies focus on AE detection and classification, such as works in the detection and classification of acoustic scenes and events (DCASE) challenges \cite{dcase2019}. 
In AEC tasks, mel spectrograms are fed into neural networks, such as CNN \cite{kong2020panns} and CRNN \cite{crnn}, to extract acoustic representations to classify audio clips. Recently, with the aid of large-scale datasets, e.g. AudioSet \cite{audioset}, and pre-trained audio models (such as PANNs \cite{kong2020panns} and AST \cite{AST}), deep learning-based models have achieved widespread success in AEC tasks. However, most AEC studies mentioned above focus only on the accuracy of AEs recognition, but have not explored the perceptual experiences that these AEs and their combinations bring to people. One of the reasons for this might be that challenges, like DCASE, do not address this aspect.
For applications like monitoring for health effects and urban planning, the perception and understanding of acoustic environments become important, and AE recognition plays a vital role \cite{boes2018machine}\cite{de2009model}. 
Emotions elicited directly by the sound or by the meaning given to it (the recognition) have multiple dimensions, but for the assessment of health effects, recent research mostly focuses on annoyance \cite{soundscape}.
To simplify their approach, researchers often only use coarse-grained audio events (cAE) labels to roughly distinguish between human, natural, and mechanical sounds. However, humans use more detailed and specific fine-grained audio events (fAE) information when understanding the sound environment \cite{de2009model}.

\vspace{-0.1cm} 
To extend AEC focusing on AEs to ARP on human perceptions, a hierarchical graph representation learning (HGRL) \cite{HGRL} has been developed recently on the DeLTA \cite{delta} dataset, the only publicly available dataset that contains both AE labels and AR. However, HGRL has shortcomings: 
1) HGRL has no corresponding semantic supervision when extracting node representations to build the graph. This leads to difficulties in specifying each node's semantics. 
2) HGRL focuses on coarse-grained global information while ignoring fine-grained local information, as it uses all nodes to build a global graph.
3) HGRL model has 92.3M parameters (\textit{params}), which may be difficult to meet requirements in some practical applications with limited resources, e.g. edge devices.
To this end, this paper proposes a lightweight attention-fused multi-level graph learning (MLGL) model with local and global semantic graphs based on nodes with explicit semantic information, which facilitates the modelling of a global graph using multiple local context-aware graphs.


The contributions are as follows: 1) Compared to HGRL, the MLGL provides higher explainability for its reliance on local and global context-aware graphs.
2) The MLGL, based on graph neural networks that capture the relations between nodes well, outperforms CNN-based models. Visual analyses show that MLGL better captures relations between multi-grained AEs, and relations between AEs and AR.
3) Statistical analysis of MLGL results shows that AEs from some sources significantly correlate with AR, which is consistent with human perception of these environmental sound sources.


\vspace{-0.1cm}
\section{Local and global context-aware MLGL}
\label{sec:format}

\vspace{-0.2cm}
In this section, we first present how to extract node representations with explicit semantic information, then use semantic nodes to build local context-aware graphs (LcGs). Next, we fuse nodes from multiple LcGs by attention mechanism to enhance common objective representations of the same node in different local contexts. Finally, we construct the global context-aware graph (GcG) with attention-enhanced nodes, which captures nodes' representation and relations from the global view. We illustrate the idea with the help of labels of 24 types of fine-grained AEs (fAEs), 7 types of coarse-grained AEs (cAEs), and human-annotated AR in DeLTA \cite{delta}.

\vspace{-0.2cm}
\subsection{Semantic node representation}\label{node_sec}

\vspace{-0.1cm} 
In Fig.~\ref{model}, MLGL uses three 3-layer VGG-like CNNs \cite{vgg} to extract representations for sound-related fAEs, cAEs, and human emotion-related AR, respectively. 
This enables MLGL to learn semantic representations of each target, which is different from HGRL~\cite{HGRL} where a shared 6-layer CNN is used to extract the mixed representations of 3 targets (fAEs, cAEs, and AR).
Then, the extracted representations are fed into the corresponding embedding block, each consisting of multiple parallel 64-dimensional embedding (64D $E_{mb}$) layers, to obtain the semantic representations for each target.  
Unlike HGRL, which has no semantic supervision for nodes in extracting their representations, MLGL ensures that the $E_{mb}$ layer only learns the semantic representation of its target, which is followed by a 1-unit linear layer and a loss function to convert its output to the target prediction. Given that the labels of fAE, cAE, and AR are $y_{f24}$, $y_{c7}$, and $y_{ar}$, respectively, the losses at level 1 are ${L}_{1}$$=$$BCE(p_{l1f24}, y_{f24})$, ${L}_{2}$$=$$MSE(p_{l1ar}, y_{ar})$, ${L}_{3}$$=$$BCE(p_{l1c7}, y_{c7})$, where BCE is binary cross entropy loss, MSE is mean squared error loss, $p_{l1f24}$, $p_{l1ar}$, and $p_{l1c7}$ are predictions of fAE, cAE, and AR at level 1, respectively. The 64D $E_{mb}$ of each target is viewed as the semantic representation of each node, which is then used to construct different LcGs at level 2.

\vspace{-0.3cm}
\subsection{Local context-aware graphs (LcGs)}
\vspace{-0.1cm}
Based on the obtained node representations, we construct 3 LcGs with different motives: 
\textbf{1) fAG}: fAEs-AR graph; 
\textbf{2) fcG}: fAEs-cAEs graph; 
\textbf{3) cAG}: cAEs-AR graph.  
Among them, the pure AE-based graph fcG aims to model the relations between the 24 classes of fAEs and higher-level cAEs, and use the contexts of AEs in different granularity to improve the node representations. 
In contrast, fAG and cAG explore the relations and complementarity between AEs and AR. Nodes in each LcG are fully connected and modelled using a 3-layer gated graph convolutional network (GCN) \cite{gated_GCN}.

\textbf{Attention-based node fusion.}
The fAG outputs 24 AR-aware fAEs (E24) node representations (marked as fAG-E24), while fcG outputs 24 cAE-aware fAEs node representations (marked as fcG-E24). Both fAG-E24 and fcG-E24 describe the same target E24 enriched by related context information in different perspectives. 
Hence, MLGL fuses the common information between these nodes by attention mechanism \cite{Transformer}.
\begin{equation}
\setlength{\abovedisplayskip}{1pt}
\setlength{\belowdisplayskip}{1pt} 
Attention(\mathbf{Q, K, V} )=softmax(\mathbf{QK^T} / \sqrt{d_{k} } )\mathbf{V} 
\label{self-attention}
\end{equation} 
where $\mathbf{V}$$=$$\mathbf{K}$, and $d_{k}$ is $\mathbf{K}$'s dimension. 
Here, $\mathbf{Q}$ acts as an index to adjust $\mathbf{V}$. The attention output is mainly based on $\mathbf{V}$, so a more informative $\mathbf{V}$ will lead to better attention results. For the E24 nodes, fAG-E24 learns E24 with 1 AR node, but fcG-E24 learns E24 with 7 cAEs nodes, so fcG-E24 will contain more context information. For fusing E24 nodes, $\mathbf{Q}$ is fAG-E24, $\mathbf{K}$ is fcG-E24, that is, using the AR-aware information to adjust the pure-AE information. For fusing the E7 nodes, $\mathbf{Q}$ is cAG-E7, and $\mathbf{K}$ is fcG-E7.
For fusing the AR1 node, $\mathbf{Q}$ is cAG-AR1, and $\mathbf{K}$ is fAG-AR1.
Similar to Section \ref{node_sec}, fused nodes are fed into the following 1-unit linear layer, and a loss is used to enhance the learning of the corresponding semantic information. 
Losses at level 2 are ${L}_{4}$$=$$BCE(p_{l2f24}, y_{f24})$, ${L}_{5}$$=$$MSE(p_{l2ar}, y_{ar})$, and ${L}_{6}$$=$$BCE(p_{l2c7}, y_{c7})$, where $p_{l2f24}$, $p_{l2ar}$, and $p_{l2c7}$ are the predictions of fAE, cAE, and AR at level 2, respectively.

\vspace{-0.3cm} 
\subsection{Global context-aware graph (GcG)}

\vspace{-0.1cm}
The GcG comprehensively considers the semantic information of fAEs, cAEs, and AR derived from multiple LcGs. 
The GcG is also modelled by a 3-layer GCN to represent its nodes and global relations between nodes. 
Global context-aware node representations are fed into the following 1-unit linear layer, and a loss is used to complete AEC and ARP tasks. 
Losses at level 3 are ${L}_{7}$$=$$BCE(p_{l3f24}, y_{f24})$, ${L}_{8}$$=$$MSE(p_{l3ar}, y_{ar})$, and ${L}_{9}$$=$$BCE(p_{l3c7}, y_{c7})$, where $p_{l3f24}$, $p_{l3ar}$, and $p_{l3c7}$ are predictions of fAE, cAE, and AR at level 3, respectively.
To perform end-to-end training for MLGL, the final loss is:
${L}$$=$$\sum_{i=1}^{9}\lambda_{i}{L}_{i}$, where $\lambda_{i}$ is the parameter for weighting losses during training, fixed to 1 in our experiments.
The losses at levels 1 and 2 enable the model to maintain semantic-related information while learning intermediate representations. However, the losses at level 3 correct the predictions of MLGL from a global perspective.

\section{Experiments and results}

\vspace{-0.2cm} 
\subsection{Dataset, experiments setup, and metrics}

\vspace{-0.1cm} 
We use the public dataset DeLTA \cite{delta} for experiments, which has 2890 15-second binaural audio clips. Each clip has labels of fAEs and the corresponding AR (continuously from 1 to 10). Following \cite{HGRL}, the labels of 7 classes cAEs are derived from the labels of 24 classes fAEs. The training, validation, and test sets contain 2200, 245, and 445 clips, respectively.

The log-mel energy with 64 banks \cite{kong2020panns} is used as acoustic features, with a Hamming window of 46 \textit{ms} with $1/3$ overlap between neighbouring windows. 
Dropout and normalization are used to prevent over-fitting of models \cite{dropout}. 
A batch size of 64 and AdamW optimizer \cite{adamw} with a learning rate of 0.0005 are used to minimize the loss. 
Models are trained on a Tesla V100 GPU for 400 epochs. Dataset, code, and models are available on the \textbf{\textit{webpage}}
{\footnotesize{(\textcolor{blue}{\underline{https://github.com/Yuanbo2020/MLGL}})}}.
Accuracy (\textit{Acc}), \textit{F-score}, and threshold-free \textit{AUC} \cite{metrics} are used to evaluate the classification results of AEC.
Mean absolute error (\textit{MAE}), mean squared error (\textit{MSE}), and R2-score (\textit{R2}) \cite{r2} are used to measure the prediction results of ARP.

\vspace{-0.3cm}
\subsection{Results and analysis}

\vspace{-0.1cm}
\textbf{Performance of different levels of prediction.}
The proposed MLGL has 9 losses at 3 levels, responsible for different roles. 
In Table \ref{tab:levels_output}, it can be observed that good predictions from all levels have been achieved for both the 24 classes of fAEs and the 7 classes of cAEs, implying that level 1 of MLGL performs well in capturing the semantic representations of base nodes.
The attention-fusion of graph-enriched representations, further improves its performance for human-perceived ARP. This implies that representations of related nodes are enhanced in level 2 graphs.
Finally, the predictions at level 3 are improved in both AEC and ARP. This reveals that node representations optimized in LcGs are beneficial for constructing a more accurate GcG.
Note that the DeLTA dataset does not have labels for coarse-grained AEC, and cAE labels from \cite{HGRL} without human validation cannot be evaluated, so later, by default, AEC will be the fine-grained AEC.

\begin{table}[H]\footnotesize
	\setlength{\abovecaptionskip}{0cm}   
	\setlength{\belowcaptionskip}{-0.2cm}  
	\renewcommand\tabcolsep{1pt} 
	\centering
	\caption{Multi-level predictions of MLGL for AEC and ARP.}
	\begin{tabular}{  
	p{0.9cm}<{\centering}|
	p{1.1cm}<{\centering}
 p{1.1cm}<{\centering}|
	p{1.1cm}<{\centering}
	p{1.1cm}<{\centering}|
 p{0.9cm}<{\centering}
 p{0.9cm}<{\centering}
 p{0.9cm}<{\centering}}
	
		\toprule[1pt] 
		\specialrule{0em}{0.1pt}{0.1pt}

 
Output &
\multicolumn{2}{c|}{Fine-grained AEC} & 
\multicolumn{2}{c|}{Coarse-grained AEC} & \multicolumn{3}{c}{ARP} \\

		\cline{2-8}     
	    Level  & \text{Acc.} (\%) & \text{AUC} & \text{Acc.} (\%) & \text{AUC} & \text{MSE} & \text{MAE} & \text{R2}  \\
	\hline 
		\specialrule{0em}{0.em}{0.pt}
  
	 1 &  91.91 &  0.918 & 84.72 &  0.910  &  1.062 & 0.767 & 0.452 \\ 
  2 &  91.90 &  0.920 & \textbf{84.98} &  0.910  &  0.964 &  0.717 & 0.502 \\
  3 &  \textbf{91.96} &  \textbf{0.921} & 84.94 &  \textbf{0.912}  & \textbf{0.940}  &  \textbf{0.706} & \textbf{0.515}\\
  
		\specialrule{0em}{0pt}{0em}
		\bottomrule[1pt]
	\end{tabular}
	\label{tab:levels_output}
\end{table}

\vspace{-0.4cm}
\noindent
\textbf{Model parameters and size.}
To simulate a realistic use case, the inference time in Table \ref{tab:computational_overhead} refers to the time from loading the raw audio clip, extracting features, feeding them into the model, and getting predictions. 
Compared with HGRL, \textit{params} of MLGL are reduced by (92.3-4.1)/92.3$\times$100\%$\approx$96\%, and the model size is reduced by about 95\%, while MLGL still improves predictions of ARP and slightly improves AEC.

\begin{table}[H]\footnotesize
	\setlength{\abovecaptionskip}{0cm}   
	\setlength{\belowcaptionskip}{-0.2cm}  
	\renewcommand\tabcolsep{1pt} 
	\centering
	\caption{Comparison of HGRL and MLGL in detail.}
	\begin{tabular}{  
	p{0.9cm}<{\centering}|
	p{0.9cm}<{\centering}|
 p{1.3cm}<{\centering}|
	p{1.1cm}<{\centering}|
 p{1.1cm}<{\centering}
	p{0.9cm}<{\centering}|
 p{0.9cm}<{\centering}
 p{0.9cm}<{\centering}}
	
		\toprule[1pt] 
		\specialrule{0em}{0.1pt}{0.1pt}  

  \multirow{2}{*}{\makecell[c]{Model}} & 
  Params & 
  Model & 
  Inference & 
  \multicolumn{2}{c|}{AEC} & 
 \multicolumn{2}{c}{ARP}    
  \\

  \cline{5-8}      
  
  & (M) & Size (MB) &  time (s) &
Acc. (\%) & AUC & MAE & R2 \\ 
\hline 
	 HGRL &  92.3 &  353.0 & 0.531 &  91.71 & 0.901  &  0.802 & 0.458 \\ 

   MLGL &  \textbf{4.1} &  \textbf{16.0} & \textbf{0.448} &   \textbf{91.96}  & \textbf{0.921}  &  \textbf{0.706} & \textbf{0.515} \\

		\specialrule{0em}{0pt}{0em}
		\bottomrule[1pt]
	\end{tabular}
	\label{tab:computational_overhead}
\end{table}


\vspace{-0.3cm}
\noindent
\textbf{Comparison of fusion methods.}
At level 2, different enriched representations of the same target node are fused by attention to enhance the common information between node representations learned from different local contexts. 
Specifically, two embedding vectors (denoted as $EV_{1}$, $EV_{2}$) of the same node are fused. Table \ref{tab:fusion_method} explores alternative fusion methods to investigate their impact on the proposed model.

\label{ssec:figure-f}
\begin{figure*}[t] 
	\setlength{\abovecaptionskip}{-0cm}  
	\setlength{\belowcaptionskip}{-0.3cm}   
	\centerline{\includegraphics[width = 1 \textwidth]{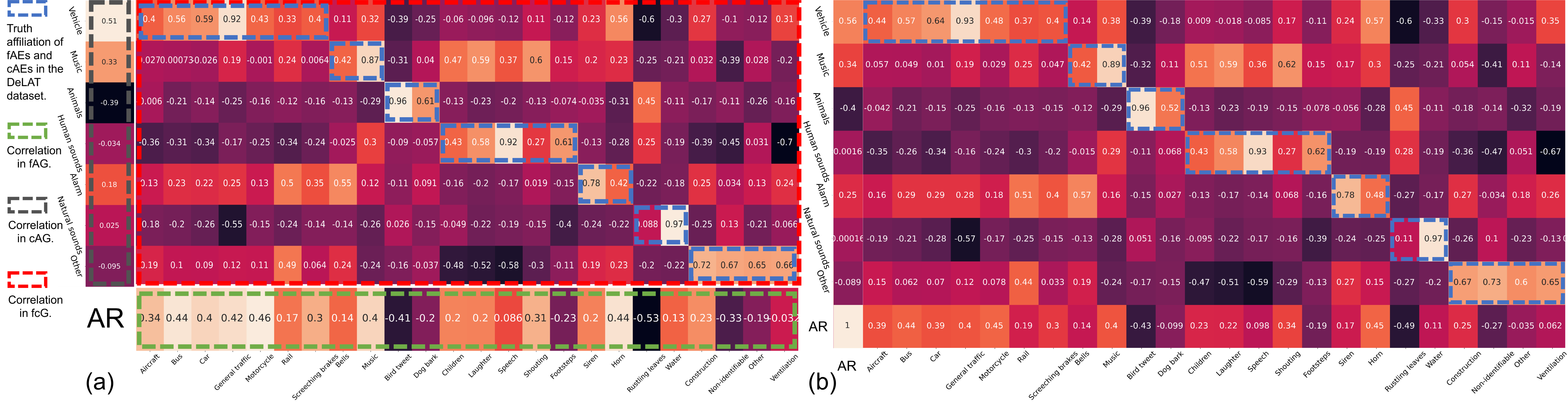}}
	\caption{(a) Correlation between node embeddings of multiple LcGs.  (b) Correlation between node embeddings of the GcG.}
	\label{event_relation1}
\end{figure*}


\begin{table}[H]\footnotesize
	\setlength{\abovecaptionskip}{0cm}   
	\setlength{\belowcaptionskip}{-0.2cm}  
	\renewcommand\tabcolsep{1pt} 
	\centering
	\caption{Results of different fusion ways at level 2 of MLGL.}
	\begin{tabular}{  
	p{0.3cm}<{\centering}|
 p{1.2cm}<{\centering}|
	p{1.2cm}<{\centering}|
 p{1.4cm}<{\centering}|
 p{1cm}<{\centering}|
 p{0.9cm}<{\centering}|
 p{1cm}<{\centering}|
 p{0.9cm}<{\centering}}
	
		\toprule[1pt] 
		\specialrule{0em}{0.1pt}{0.1pt} 
 
 \multirow{2}{*}{\makecell[c]{\#}} & Fusion &
\multicolumn{3}{c|}{AEC} & 
 \multicolumn{3}{c}{ARP} \\

		\cline{3-8}     
	 & Type  & \text{Acc.} (\%) & \text{F-score} (\%)  & \text{AUC} & \text{MSE} & \text{MAE} & \text{R2}  \\
	\hline 
		\specialrule{0em}{0.em}{0.pt}
 
	 1 & Addition & 91.67 & 67.96 & 0.897 & 1.247 & 0.846 & 0.356 \\
  2 & Concat & 91.68 & \textbf{68.67} & 0.903 & 1.148 & 0.818 & 0.407 \\
  3 & Hadamard & 91.73 & 68.58 & 0.908 & 1.282 & 0.872 & 0.339 \\
  4 & Gating & 91.75 & 68.11 & 0.907 & 1.141 & 0.813 & 0.411 \\
  5 & Attention & \textbf{91.96} & 68.36 & \textbf{0.921} & \textbf{0.940} & \textbf{0.706} & \textbf{0.515}\\

		\specialrule{0em}{0pt}{0em}
		\bottomrule[1pt]
	\end{tabular}
	\label{tab:fusion_method}
\end{table}

\vspace{-0.3cm}
In Table \ref{tab:fusion_method}, \#1 refers to the addition of two vectors, while \#2 refers to the concatenation of two vectors, which is then fed to a linear layer. Therefore, in \#2, an extra layer is added with respect to \#1. \#3 performs the Hadamard product of two vectors. Similar to gated linear units \cite{GLU}, \#4 utilizes the gated output of $EV_{2}$ to condition $EV_{1}$, $(W_{1}\times EV_{1} +b_{1}) \odot \sigma (W_{2}\times EV_{2} + b_{1})$, where $\sigma$ is a sigmoid function, $\odot$ is element-wise product, $W$ and $b$ are learnable weight and bias, respectively.
Overall, the AEC performance of these fusion methods is similar, while attention-based fusion improves the score on threshold-free AUC. For ARP, the attention-based fusion significantly improved the scores on MSE, MAE, and R2. Please see the \textbf{\textit{webpage}} for source code and models.


\noindent
\textbf{Comparison to other methods.}
We compare the performance of models in Table \ref{tab:other_models}, where the results of CNN, CNN-Transformer, PANNs, and HGRL are all taken from \cite{HGRL}. Compared with these models, MLGL, despite using much fewer parameters, achieves better results in both AEC and ARP, with MSE less than 1, and R2 greater than 0.5 for ARP.

\begin{table}[H] \footnotesize 
\setlength{\abovecaptionskip}{0cm}   
	\setlength{\belowcaptionskip}{-0.2cm}
	\renewcommand\tabcolsep{1pt} 
	\centering
	\caption{Comparison of different models on DeLTA dataset.}
	\begin{tabular}{
	p{0.3cm}<{\centering}| 
 p{1.5cm}<{\centering}| 
 p{1cm}<{\centering} |
	p{1.3cm}<{\centering}
 p{1cm}<{\centering}|
 p{0.9cm}<{\centering}
 p{1cm}<{\centering}
 p{0.9cm}<{\centering}
	} 
	   \toprule[1pt] 
    \specialrule{0em}{0.1pt}{0.1pt} 
    
		\multirow{2}{*}{\makecell[c]{\#}} & \multirow{2}{*}{\makecell[c]{Model}} &  Param. &
\multicolumn{2}{c|}{ AEC} & \multicolumn{3}{c}{ ARP} \\

    \cline{4-8}
  & & (M) & \text{F-score}(\%) & AUC & \text{MSE} & \text{MAE} & \text{R2}  \\
  
		\hline

  1 & CNN  & 0.8 & 55.05 &  0.891 & 1.675 & 0.997 & 0.135 \\

  2 & CNN-Trans.  & 1.6 & 58.66  &  0.851 & 1.445 & 0.966 & 0.254  \\
 
 3 & PANNs & 79.7 & 63.86 &  0.903 & 1.162  & 0.858 & 0.400 \\

 4 & HGRL & 92.3 & 67.91 &  0.901 &  \text{1.049} & \text{0.802} & \text{0.458} \\

 5 & MLGL & 4.1 & \textbf{68.36} &  \textbf{0.921} &  \textbf{0.940} & \textbf{0.706} & \textbf{0.515} \\
	 
	\specialrule{0em}{0pt}{0em}
		\bottomrule[1pt]
	\end{tabular}
	\label{tab:other_models}
\end{table}

\vspace{-0.2cm}
\noindent
\textbf{Explaining how the model works.}
To study the ability of LcGs to capture relations between nodes, and whether the GcG can model three kinds of semantic nodes in one graph, Fig. \ref{event_relation1} visualizes the Pearson correlation coefficients \cite{pcc} between node embeddings on all the test audio clips. 
In Fig. \ref{event_relation1}, the correlations between the nodes in fcG successfully match the associations of fAEs to cAEs in the DeLTA dataset. The correlations in the fAG indicate that \textit{motorcycle}, \textit{bus}, and \textit{horn} are the most likely sounds to cause people annoyance, while \textit{rustling leaves} and \textit{bird tweet} sounds are the least likely to be annoying.
The cAG box shows that \textit{vehicles} sounds are more likely to annoy people, while \textit{animals} sounds are not.
These LcGs with separate motivations perform well in capturing the semantic relations between internal nodes.

In Fig. \ref{event_relation1} (b), the GcG aligns all the semantic nodes using one graph. HGRL \cite{HGRL} has similar results, but the correlation of nodes in Fig.~\ref{event_relation1} (b) is significantly higher than that of HGRL. This means that MLGL can model different semantic nodes and their relations from a global view, resulting in better performance on AEC and ARP tasks, as compared with HGRL, because MLGL has nodes with explicit semantics and uses local context-aware graphs to optimize the node representations. For a clearer Fig. \ref{event_relation1}, please see the \textbf{\textit{webpage}}.


\noindent
\textbf{Statistical significance of correlations.}
For the results in Fig. \ref{event_relation1}, Table \ref{tab:spearman} further tests their significance. Due to limited space, Table \ref{tab:spearman} shows partial AEs. We first assess the distribution of MLGL's predictions with a Shapiro–Wilk test. Since the predictions are not normally distributed, Spearman's rank (Spearman's rho) \cite{rho} correlation coefficient (CC) is used. Note: ** indicates statistical significance at the 0.001 level.

\begin{table}[H]\footnotesize
	\setlength{\abovecaptionskip}{0cm}   
	\setlength{\belowcaptionskip}{-0.2cm}  
	\renewcommand\tabcolsep{1pt} 
	\centering
	\caption{Spearman’s rho correlation coefficients for AEs with AR.}
	\begin{tabular}{  
	p{0.2cm}<{\centering}|
 p{0.65cm}<{\centering}|
	p{1.2cm}<{\centering}|
 p{0.3cm}<{\centering}|
 p{1.25cm}<{\centering}|
	p{1.2cm}<{\centering}|
 p{0.3cm}<{\centering}|
 p{1.75cm}<{\centering}|
	p{1.2cm}<{\centering}}
	
		\toprule[1pt] 
		\specialrule{0em}{0.1pt}{0.1pt}  

  \# & 
  AE & 
  CC w/ AR &
  \# & 
  AE & 
  CC w/ AR&
  \# & 
  AE & 
  CC w/ AR
  \\

\hline 

1& 	Bus	& 0.504**   & 7	& Speech	& 0.193** & 13 & 	General traffic& 	0.391** \\
2	& Car& 	0.199**   & 8	& Children	& 0.173**  & 14	& Motorcycle& 	0.374** \\
3	& Rail	& 0.329** & 9& Shouting	& 0.443** & 15 & 	Screech brakes& 	0.488** \\
4& 	Bells	& 0.458**   & 10 & Bird tweet	& -0.522** & 16	& Rustling leaves	& -0.585** \\
5	& Horn	& 0.653** & 11	& Dog bark	& 0.029** & 17	& Ventilation	& 0.085 \\
6	& Water	& -0.097** & 12 & Aircraft & 0.415** & 18	& Other	& 0.011 \\
 
		\specialrule{0em}{0pt}{0em}
		\bottomrule[1pt]
	\end{tabular}
	\label{tab:spearman}
\end{table}

\vspace{-0.3cm}
In Table \ref{tab:spearman}, several AEs, namely \textit{Aircraft}, \textit{Bus}, \textit{Screeching brakes}, \textit{Bells}, \textit{Shouting}, and \textit{Horn} exhibit strong positive correlations with AR, which indicates that an increase in the occurrences of these AEs increases the level of annoyance. In contrast, AEs like \textit{Bird tweet} and \textit{Rustling leaves} show strong negative correlations with AR, implying a decrease in annoyance when these are present. These statistical results from the proposed MLGL are consistent with earlier human-perception-based soundscape research \cite{de2009model}\cite{boes2018machine}.


\vspace{-0.2cm}
\section{CONCLUSION}
\label{sec:CONCLUSION}

\vspace{-0.1cm} 
We have presented the MLGL to identify audio events (AEs) generated by diverse environmental 
sound sources and predict human-perceived annoyance rating (AR) in real-life soundscapes. Experimental results show that: 1) Lightweight MLGL with only 4.1 M parameters performs AEC and ARP tasks as well as previous heavier models;  
2) Local context-aware graphs with different motivations perform well in capturing semantic nodes, which helps the global-view graph to model the nodes and their relations; 
3) Statistical analysis of MLGL results shows that some AEs are closely related to human AR, which is consistent with observations from previous soundscape research based on human perception. Future work will deploy MLGL on edge devices and measure differences in real applications to assist soundscape research.

 \vspace{-0.1cm}
 \section{ACKNOWLEDGEMENTS}
\label{sec:ACKNOWLEDGEMENTS}

 \vspace{-0.2cm}
This research received funding from the Flemish Government under the “Onderzoeksprogramma Artificiële Intelligentie (AI) Vlaanderen” programme.

\vfill\pagebreak

\label{sec:refs}

\bibliographystyle{IEEEbib}
\bibliography{Template}

\end{document}